\documentclass[11pt,twoside]{article}

\usepackage{asp2004}
\usepackage{epsf}
\usepackage{psfig}
\usepackage{lscape}
\usepackage{natbib}

\begin{document}
\title{A 15 $\micron$ selected sample of high-z starbursts and AGNs}  
\author{A. Hern\'an-Caballero$^{1}$, I. P\'erez-Fournon$^{1}$, M. Rowan-Robinson$^{2}$, D. Rigopoulou$^{3}$,
        A. Afonso-Luis$^{1}$, E. Hatziminaoglou$^{1}$, E. Gonz\'alez-Solares$^{4}$, F.M. Montenegro-Montes$^{5}$,
        B. Vila-Vilaro$^{6}$, D. Farrah$^{7}$, C. Lari$^{5}$, M. Vaccari$^{2}$, T. Babbedge$^{2}$, S. Oliver$^{8}$,
        D. Clements$^{2}$, S. Serjeant$^{9}$, F. Pozzi$^{10}$, F. La Franca$^{11}$, C. Gruppioni$^{10}$,
        I. Valtchanov$^{2}$, C. Lonsdale$^{7}$ and the SWIRE team}  
\affil{$^{1}$Instituto de Astrof\'isica de Canarias, La Laguna, Tenerife (Spain), 
$^{2}$Imperial College, London (UK),
$^{3}$Oxford University (UK),
$^{4}$Institute of Astronomy, University of Cambridge (UK),
$^{5}$Istituto di Radioastronomia, Bologna (Italy),
$^{6}$NAOJ, Tokyo (Japan),
$^{7}$Cornell University (USA),
$^{8}$University of Sussex (UK),
$^{9}$University of Kent (UK),
$^{10}$University of Bologna (Italy),
$^{11}$Universita degli Studi ``Roma Tre'' (Italy)
} 

\begin{abstract} 
We report results from our Spitzer GO-1 program on IRS spectroscopy of
a large sample of Luminous Infrared Galaxies and quasars selected from
the European Large Area ISO Survey (ELAIS). The selected ELAIS sources 
have a wide multi-wavelength
coverage, including ISOCAM, ISOPHOT, IRAC and MIPS (from SWIRE), and optical
photometry. Here we present the sample selection and results from the IRS 
spectroscopy.
\end{abstract}

\section{Sample Selection and IRS Observations}

The sources were selected from the European Large Area ISO Survey
(ELAIS) final band-merged catalog of \citet{rrw04}. 
The sample consists of 70 sources with 15 $\micron$ fluxes larger
than $\sim$ 1 mJy and spectroscopic or 
estimated photometric redshifts $z > 1$.
Although no color cuts were applied, the
objects are brighter than $r$ $\sim$ 24, the limit of the Isaac
Newton Telescope Wide Field Survey CCD photometry used in the
optical identification of ELAIS sources \citep{gonzalez05}.
The 15 $\micron$ observations and catalog are presented in
\citet{vaccari05}.\\

Low-resolution IRS spectroscopy was carried out
using all four IRS modules, covering thus the wavelength range 
between 5 and 40 $\micron$. 
Typical total exposure time per object was of about one hour.
Figure 1 shows the ratio of
$\nu$\textit{f}$_\nu$ at 15 $\micron$ over $r$-band
as a function of the $r$-band magnitude. ELAIS-IRS targets
with bright magnitudes have blue optical to 15 $\micron$ colors
typical of type-1 AGN \citep{afonso04,gonzalez05,hatzim05}.
Objects with fainter optical IDs are identified as obscured
AGN and star-forming galaxies.

\begin{figure}[tb]
\plotone{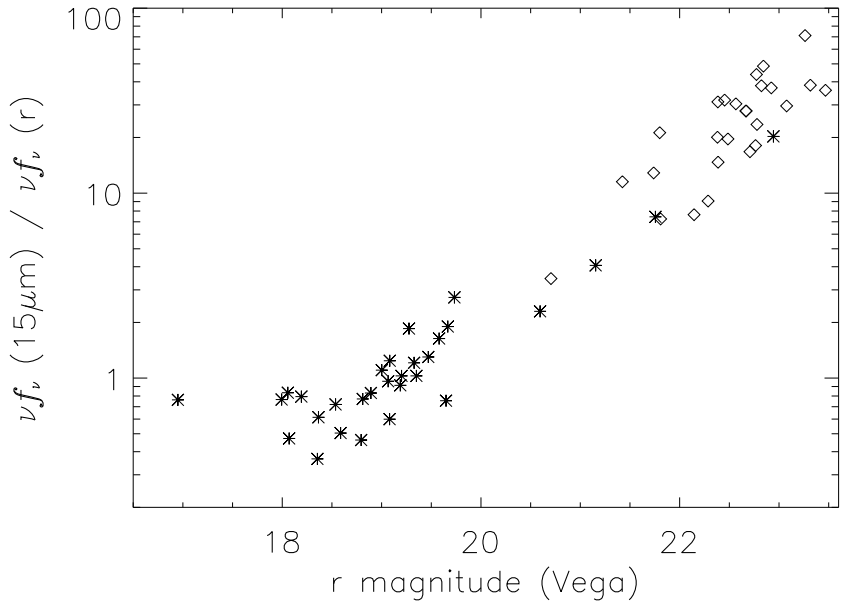}
\label{ratio}
\caption{Ratio $\nu \textit{f}_\nu$ (15 $\micron$) / 
$\nu \textit{f}_\nu$ ($r$)
for the ELAIS-IRS sources versus $r$-band magnitude. Diamonds: star-forming
galaxies and obscured AGN; stars: unobscured AGN.}
\end{figure}

\section{Results from the IRS Spectroscopy}

The IRS spectra were extracted from the SSC pipeline processed
data using SPICE, and individual spectra of each object were 
coadded. A selection of the IRS spectra is shown in figure
2. The IRS spectra show a wide variety
of spectral shapes and clear features (PAHs in emission and silicate
absorption at 9.7 $\micron$) can be seen in a number of objects.
The IRS spectra can be classified into three main categories:
(a) smooth featureless continuum, usually associated with type-1 AGN,
(b) PAH features in emission and silicate absorption, and
(c) silicate absorption.

\begin{figure}[tb]
\plotone{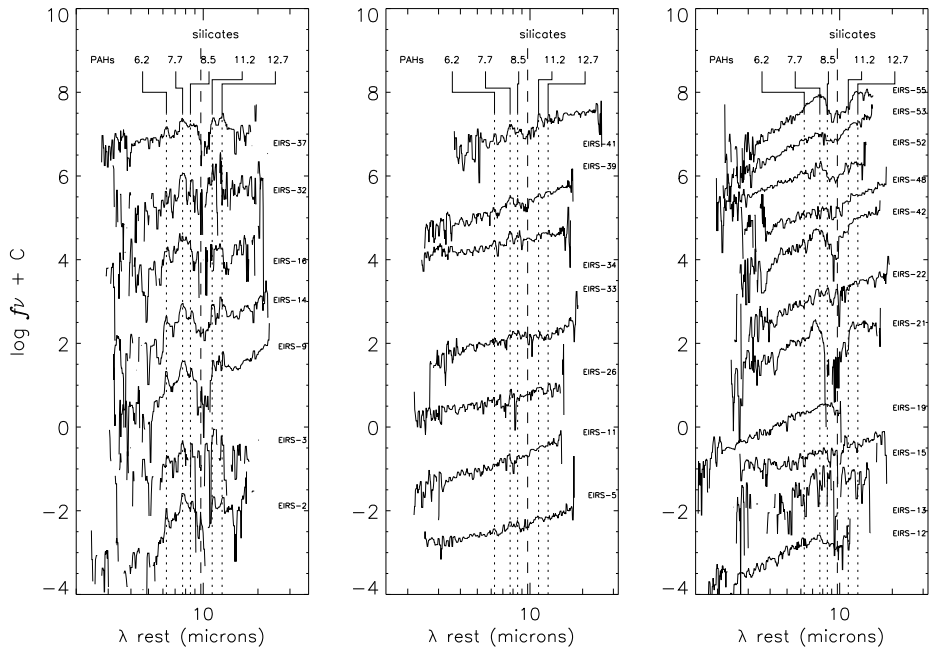}
\label{spectra}
\caption{A selection of ELAIS-IRS spectra showing different types of mid-IR
SEDs, including strong PAH, star-forming dominated SEDs (left), power-law continua
with or without PAH features (center) and silicate absorption (right).}
\end{figure}

\begin{figure}[tb]
\plotone{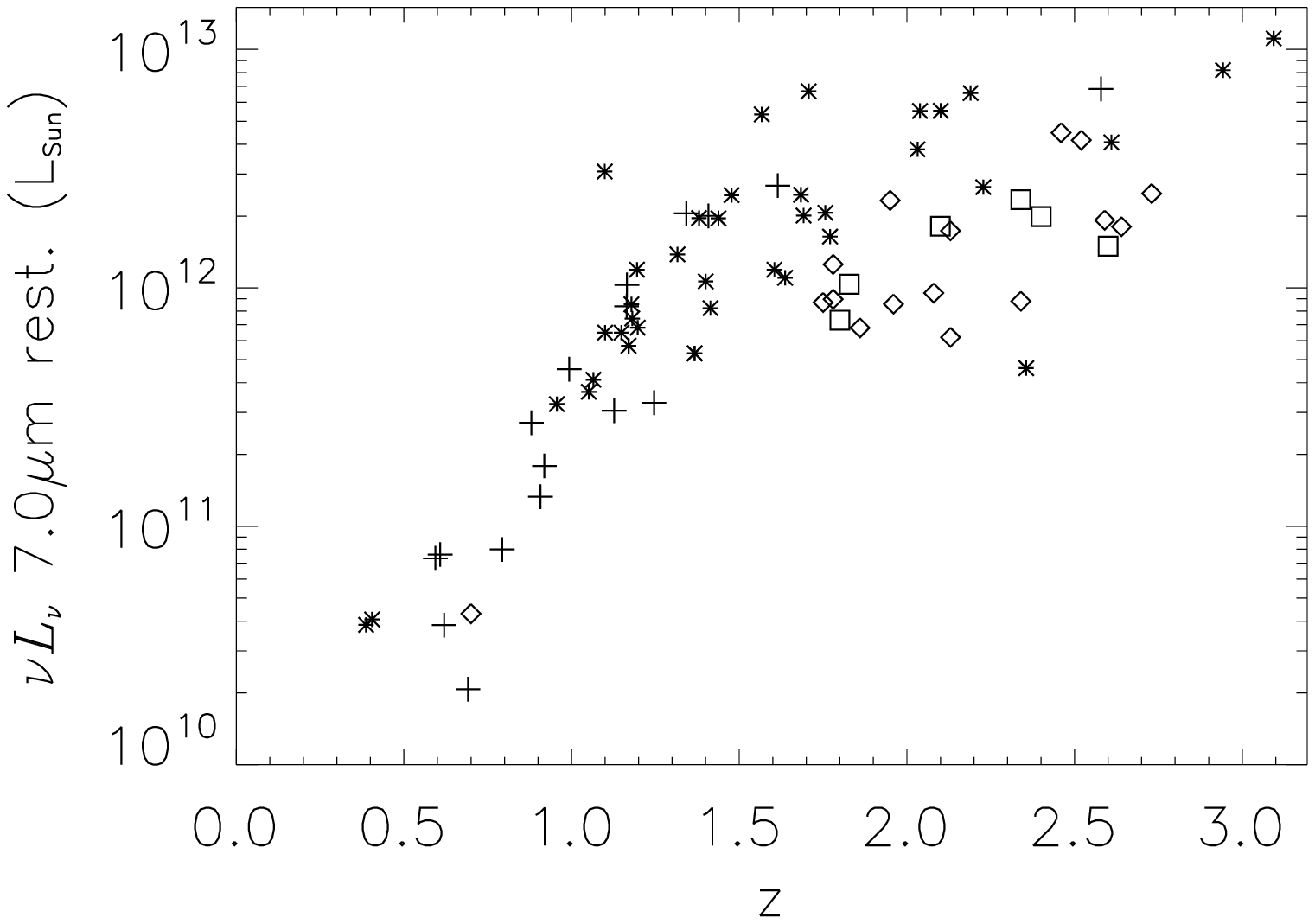}
\label{nuLnu}
\caption{Restframe 7 $\micron$ luminosity ($\nu \textit{L}_\nu$) versus redshift
for the ELAIS-IRS sample. Objects with optical spectroscopic 
redshifts (mostly AGN) are shown as stars, whereas those with $z$ estimated from
their IRS spectra are plotted as plus signs. For comparison, samples from
\citet{yan05} and \citet{houck05} are represented as squares and diamonds,
respectively.}
\end{figure}

Redshifts can be measured for a number of objects from the IRS
spectroscopy, and they agree with the optical spectroscopic
redshifts, whenever available. They are also consistent in most cases with the 
photometric redshifts obtained from template fitting to the optical
and IRAC photometry from SWIRE \citep{lonsdale03,lonsdale04}.
Objects in our sample with PAH features and starlight-dominated
optical SEDs are interpreted as star-forming galaxies. Their redshifts
are in the range 0.6 $\la$ z $\la$ 1.2 and the far-IR (8-1000 $\micron$) 
luminosities are in the range $\sim$ $10^{11}$ - $10^{12}$ $L_{\odot}$.\\

Some galaxies are found at higher redshifts, up to z $\sim$ 2. They 
only show silicate in absorption and no bright PAH features. Their
luminosities are in the ULIRG range. Their SEDs and IRS spectra
suggest they are obscured AGN.

\section{Comparison with other IRS samples of high-z galaxies}

Our sample, selected at 15 $\micron$ from the ELAIS survey, differs
in redshift range and luminosity from those in other major IRS surveys (fig.
3).
The sample selection of \citet{yan05} was based in color cuts using the
Spitzer 24 and 8 $\micron$ bands and one optical band (R), while
\citet{houck05} selected objects with very red 24 $\micron$ to R band 
colors. Our ELAIS-IRS sample selection aims to cover all possible types
of z $\ga$ 1 sources selected at 15 $\micron$, regardless of their mid-IR
colors. The results show that a large fraction of the star-forming galaxies
in the sample is at z $\sim$ 1, as expected from the photometric redshifts
estimates, and is consistent with bright PAH features (7.7 and 8.5 $\micron$)
redshifted into the ISOCAM LW3 band (15 $\micron$).
At low redshift (z $\la$ 1) our sample includes luminous infrared star-forming 
galaxies and two AGN at intermediate redshifts (1.0 $\la$ z $\la$ 1.8) we find 
AGN, both obscured and unobscured, and star-forming galaxies, all with 
restframe 7 $\micron$ luminosities comparable to those in the \citet{yan05} 
and \citet{houck05} samples. At z $\ga$ 1.8 we find  AGNs, typically 
more luminous than the comparison IRS samples.\\ 
 
The ELAIS-IRS sample constitutes one of the best samples of luminous and 
ultraluminous IR sources with IRS spectroscopy in the redshift range 
(0.5 $\la$ z $\la$ 1.8). A number of spectroscopic follow-up programs are underway. 
The ELAIS-IRS sample and detailed results are presented in \citet{hernan06}.\\

\acknowledgements 

This work is based on observations made with the \textit{Spitzer Space
Telescope}, which is operated by the Jet Propulsion Laboratory, Caltech
under NASA contract 1407.

\end{document}